\documentclass[twocolumn]{mn}
\usepackage {graphicx,subfigure,times}
\input{psfig}
\def\sigs{\sigma_{\rm star}}
\def\sigg{\sigma_{\rm gas}}
\def\Sigs{\Sigma_{\rm star}}
\def\Sigg{\Sigma_{\rm gas}}

\def\Qs{Q_{\rm star}}
\def\Qg{Q_{\rm gas}}
\def\delg{\delta_{{\rm gas}}}
\def\dels{\delta_{{\rm star}}}
\def\deld{\delta_{{\rm disc}}}
\def\Mdg{M_{\rm gas, disc}}
\def\Mds{M_{\rm star, disc}}
\def\Mgacc{\dot{M}_{\rm gas, acc}}
\def\Msacc{\dot{M}_{\rm star, acc}}
\def\Mgev{\dot{M}_{\rm gas,inflow}}
\def\Msev{\dot{M}_{\rm star,inflow}}
\def\Md{M_{\rm disc}}
\def\Mt{M_{\rm tot}}
\def\Mb{M_{\rm bar}}

\def\Rd{R_{\rm disc}}

\def\gamdis{\gamma_{\rm dis}}
\def\gamacc{\gamma_{\rm acc}}
\def\gamout{\gamma_{\rm out}}
\def\epssfr{\epsilon_{\rm sfr}}
\def\Sig85{\Sigma_{\rm gas,85}}
\def\Q2c{Q_{\rm 2c}}

\newcommand{\be}{\begin{equation}}
\newcommand{\ee}{\end{equation}}
\newcommand{\bea}{\begin{eqnarray}}
\newcommand{\eea}{\end{eqnarray}}
\newcommand{\equ}[1]{eq.~(\ref{eq:#1})}

\newcommand{\Equ}[1]{Eq.~(\ref{eq:#1})}

\newcommand{\se}[1]{\S\ref{sec:#1}}
\newcommand{\fig}[1]{Fig.~\ref{fig:#1}}

\newcommand{\Fig}[1]{Figure~\ref{fig:#1}}

\newcommand{\no}{\noindent}

\newcommand{\msun}{M_\odot}
\newcommand{\Msun}{M_\odot}

\newcommand{\ifm}[1]{\relax\ifmmode#1\else$\mathsurround=0pt #1$\fi}
\newcommand{\kms}{\ifmmode\,{\rm km}\,{\rm s}^{-1}\else km$\,$s$^{-1}$\fi}

\newcommand{\Mpc}{\,{\rm Mpc}}
\newcommand{\kpc}{\,{\rm kpc}}
\newcommand{\pc}{\,{\rm pc}}
\newcommand{\Gyr}{\,{\rm Gyr}}

\newcommand{\ltsima}{$\; \buildrel < \over \sim \;$}
\newcommand{\lsim}{\lower.5ex\hbox{\ltsima}}
\newcommand{\gtsima}{$\; \buildrel > \over \sim \;$}
\newcommand{\gsim}{\lower.5ex\hbox{\gtsima}}

\newcommand{\const}{\rm const.}

\def\la{\langle}

\begin{document}

\title[Two-Component Instability in Galactic discs]
{
Evolution of violent gravitational disc instability in galaxies: \\
Late stabilization by transition from gas to stellar dominance
}

\author[Cacciato, Dekel \& Genel]
       {\parbox[t]{\textwidth}{
        Marcello Cacciato$^{1}$\thanks{Minerva Fellow 
           \newline E-mail: cacciato@phys.huji.ac.il}, 
        Avishai Dekel$^1$
        \& 
        Shy Genel$^{2,3}$}\\ 
           \vspace*{3pt} \\
	$^1$Racah Institute of Physics, The Hebrew University, Jerusalem 91904, Israel\\
	$^2$Max-Planck-Institut f\"ur extraterrestrische Physik, Giessenbachstrasse, D-85748 Garching, Germany\\
	$^3$School of Physics and Astronomy, Tel Aviv University, Tel Aviv 69978, Israel}

\date{}

\pagerange{\pageref{firstpage}--\pageref{lastpage}}
\pubyear{2008}

\maketitle

\label{firstpage}

\begin{abstract}
We address the cosmological evolution of violent gravitational instability
in high-redshift, massive, star-forming galactic discs.
To this aim, we integrate in time the equations of mass and energy conservation 
under self-regulated instability of a two-component disc of gas and stars.
The disc is assumed to be continuously fed by cold gas at the average 
cosmological rate.
The gas forms stars and is partly driven away by stellar feedback.
The gas and stars flow inward through the disc to a central bulge 
due to torques that drive angular momentum outwards.
The gravitational energy released by the mass inflow down the gravitational 
potential gradient drives the disc turbulence that maintains the disc 
unstable with a Toomre instability parameter $Q \sim 1$,
compensating for the dissipative losses of the gas turbulence and 
raising the stellar velocity dispersion. 
We follow the velocity dispersion of stars and gas as they heat and cool
respectively and search for disc ``stabilization", 
to be marked by a low gas velocity dispersion 
comparable to the speed of sound $\sim 10 \kms$.
We vary the model parameters that characterize the
accreted gas fraction, turbulence dissipation rate, star-formation rate,
and stellar feedback.  
We find that as long as the gas input roughly follows the average cosmological
rate, the disc instability is a robust phenomenon at high redshift till 
$z \sim 1$, driven by the high surface density and high gas fraction due
to the intense cosmological accretion.
For a broad range of model parameter values, the discs tend to 
``stabilize" at $z \sim 0-0.5$ as they become dominated by hot stars.
When the model parameters are pushed to extreme values, the discs
may stabilize as early as $z \sim 2$, with the gas loss by strong outflows
serving as the dominant stabilizing factor.
\end{abstract}


\section{Introduction}
\label{sec:intro}

According to our current understanding, high-redshift massive
galaxies form in virialized dark-matter haloes that reside 
at the nodes of the cosmic web. 
When baryons are accreted into a halo along the filaments of this 
web (Dekel et al. 2009), 
angular-momentum conservation implies that most of the baryons settle 
into a disc, rotating with a circular velocity 
that is roughly comparable to the virial velocity of the halo 
(Fall \& Efstathiou 1980; Mo et al. 1998; Bullock et al 2001; but see
Danovich et al. 2011).
If the disc is gravitationally unstable, with a Toomre parameter $Q \sim 1$ 
(Toomre 1964, see below), the velocity dispersion $\sigma$ can
be estimated from the circular velocity and the mass fraction in the cold disc 
component (e.g. Dekel, Sari \& Ceverino 2009).

Pioneering observations of massive galaxies at $z \sim 2$ 
have revealed striking differences between
local and high-redshift star-forming discs of comparable masses and sizes
(Elmegreen \& Elmegreen 2005; Genzel et al. 2008).
In particular, the high-redshift discs of $\sim 10^{11}\msun$ extending to
$\sim 10\kpc$ are thick, perturbed and highly 
turbulent, with $\sigma \sim 30-80\kms$, compared to the thin, more uniform
local discs with $\sigma \sim 10 \kms$.
The high-redshift discs are gravitationally unstable, showing large transient
perturbations and bound clumps of $\sim 10^9\msun$ and sizes $\sim 1\kpc$,
as opposed to the more uniform mass distribution in typical local discs,
in which the molecular clouds are smaller by a few orders of magnitude.
While all discs may be gravitationally unstable to some degree,
the instability in the high-redshift discs is characterized by higher velocity
dispersion and more massive perturbations and thus by a more rapid dynamical 
evolution, such as inward migration on an orbital timescale, which we term
``violent gravitational disc instability". This is as opposed to the secular
evolution associated with spiral arms and bars in low-redshift discs.

The violent instability of high redshift discs can be explained in the 
framework of a steady intense gas supply from the cosmic web 
and a rapid mass migration due to gravitational instability in the disc. 
The gravitational fragmentation of gas-rich, thick, turbulent discs into clumps
and the subsequent migration into a central bulge have been proposed (van den 
Bergh et al. 1996; Elmegreen \& Elmegreen 2005; Genzel et al. 2008; Bournaud et
al. 2008) and successfully simulated for idealized discs in isolation (Noguchi
1999; Immeli et al. 2004; Bournaud et al. 2007). Violent gravitational disc
instability in the full cosmological context has been simulated 
by Agertz et al. (2008) and Ceverino, Dekel \& Bournaud (2010),
demonstrating self-regulated instability in steady state for several Gyr. 
Dekel, Sari \& Ceverino (2009, hereafter DSC09) applied a Toomre instability
analysis (Toomre 1964) to high-redshift discs under the assumption that
they are made of one cold component and fed at the average cosmological 
accretion rate. This analysis led to the tentative conclusion that an 
unstable disc at high redshift remains unstable in a cosmological steady 
state. This fails to account for the evolution from violently unstable 
discs at high redshift to marginally unstable or stable discs at 
low redshift, as indicated by observations and numerical simulations 
(e.g. \S~9 in Ceverino et al. 2011; Martig et al. 2009). 

In this paper, we generalize the DSC09 analysis by allowing the gas to 
continuously form stars. We apply a two-component disc instability analysis 
(Jog \& Solomon 1984; Rafikov 2001; Wang \& Silk 1994; Romeo \& Wiegert 2011).
The disc is assumed to be fed by fresh gas at a constant fraction of the 
average cosmological accretion rate and the
gravitational instability in the disc induces mass inflow to a central bulge.
The gravitational energy released by this inflow is assumed to drive 
velocity dispersion in the two components of the disc. 
With time, the dissipationless stellar
component acquires high velocity dispersion, whereas the gas turbulence 
dissipates its energy on a dynamical tiumescale.
We solve for the velocity dispersions of the two components under conservation
of mass and energy and the assumption that the disc instability
is self-regulated at a $Q \sim 1$ state. 
At late times, as the accretion rate decreases and the disc becomes
dominated by the hot stellar component, the gas is required to 
have a lower velocity dispersion in order to maintain the instability.
When the required gas velocity dispersion becomes comparable to the thermal
speed of sound, $c_{\rm s} \sim 10 \kms$, the pressure cannot keep decreasing 
to the level where gravitational instability is possible, and we associate
it with the end of the instability phase.
We examine whether the dominance of the stellar component
and the energy balance associated with the dissipation of gas turbulence
and inflow down the potential gradient can lead to
stabilization before $z=0$.

This paper is organized as follows.
In \S~\ref{sec:model} we introduce the ingredients of the model that describes
the evolution of a self-regulated disc instability in a cosmological context. 
In \S~3 we re-visit the one-component case, 
separately for stars or gas,
and introduce the fiducial model for a two-component
disc of gas and stars.
In \S~4 we address the two-componet evolution
for values of the model parameters in a plausible range. 
In \S~5 we discuss our results and draw conclusions.

Throughout the paper, we assume a flat $\Lambda$CDM cosmology 
specified by the cosmological parameters 
$(\Omega_{\rm m}, \sigma_8, n, h)=(0.27, 0.81, 0.96, 0.70)$,
motivated by the WMAP7+BAO+H0 results (Komatsu et al. 2011). 
The Hubble time in Gyr at redshift $z$ is denoted by $t_{\rm H}(z)$.

\section{The Model}
\label{sec:model}
We integrate in time the equations of mass and energy
conservation under self-regulated, two-component disc instability. 
Before embarking on the details of the model, we present its three
key components.

As a galactic disc of gas mass $M_{\rm gas}$ evolves, it obeys a simple 
{\it mass conservation\,} equation of the form
\be
\dot{M}_{\rm gas} = \dot{M}_{\rm source} - \dot{M}_{\rm sink} \, ,
\label{eq:mass}
\ee
where ${\dot M}_{\rm source}$ is a given gas accretion rate
of cosmological origin (\se{cosmoaccr}), 
and the sink term consists of gas conversion into stars, gas 
inflow into the disc centre and 
possible mass expulsions in outflows. 
We assume that the sink term has the general form
\be
\label{eq:mdotsink}
\dot{M}_{\rm sink}=M_{\rm gas}/\tau \, ,
\ee
where $\tau$ is a characteristic timescale.
If $\dot{M}_{\rm source}$ and $\tau$ vary sufficiently slowly in time,
then the equation converges on a timescale $\tau$ ($\propto e^{-t/\tau}$) 
to a steady-state solution where  
$\dot{M}_{\rm gas} =0$, 
$\dot{M}_{\rm source} = \dot{M}_{\rm sink}$, 
and $M_{\rm gas} = {\dot M}_{\rm source}\,\tau$.
Analytic models based on variants of \equ{mdotsink} have been attempted
for various purposes. Bouch\`e et al. (2010) and Krumholz \& Dekel (2011) 
considered draining by star formation but ignored migration.
On the other hand, DSC09 considereed migration but ignored star formation and
outflows. 
Dav\`e et al. (2011) considered star formation and feedback
as well as the return of outflowing gas back to the disc.
Here, we consider the three sink terms of
mass inflow towards the galaxy centre, 
star formation, and gas outflows due to stellar feedback.

A second important component in our model is the explicit appeal
to {\it energy conservation} in the context of a self-regulated disc
instability (\se{energy}). 
This is a new element in comparison to 
earlier studies (DSC09; Bouch\`e et al . 2010) and it is in line with the 
approach of Krumholz and Burkert (2010).
At high redshift, galactic discs are expected to have a high surface 
density, reflecting the high mean cosmological density. 
The high gas accretion rate
results in a high gas fraction, especially at $t < \tau$, when the sink term
cannot yet catch up with the source term.  
Together, they lead to gravitational instability.
In the perturbed discs, gravitational torques drive angular momentum out 
and thus cause mass inflow towards the centre, $\dot{M}_{\rm inflow}$,
(Gammie et al. 2002; DSC09; Krumholz and Burkert 2011;
 Bournaud et al. 2011).
This inflow down the potential well between the disc outskirts and its centre, 
which is on the order of the
circular velocity squared, $V^2_{\rm circ}$, provides an energy gain of
${\dot M}_{\rm inflow} V^2_{\rm circ}$, which can be used to stir up 
turbulence in the gas, characterized by a velocity dispersion $\sigg$. 
Assuming that this energy gain roughly compensates 
for the dissipative losses of the turbulence, one can write
\be
\label{eq:approxenergy}
{\dot M}_{\rm inflow} V^2_{\rm circ} 
\simeq \frac{M_{\rm gas} \sigma^2_{\rm gas}}{t_{\rm dis}} \ .
\ee  
Here, $t_{\rm dis}$ is the timescale over which the internal turbulent
energy of the gas is dissipated. It is expected to be 
comparable to one or a few disc dynamical times, 
defined as $t_{\rm dyn} = \Omega^{-1} = \Rd/V_{\rm circ}$,
where $\Rd$ and $\Omega$ are the effective disc radius and angular velocity.
The dissipation rate thus determines the disc inflow rate that is needed for self-regulating the disc instability at $Q \sim 1$.

The third key and new element of the model is the enforcement of self-regulated
instability in a {\it two-component\,} disc made of gas and stars 
(\se{instability}). 
A disc is unstable once the self-gravitational attraction acting on a patch
(associated with the surface density of cold material $\Sigma$) 
overcomes the reppeling forces due to pressure (asociated with $\sigma$) 
and rotation (associated with $\Omega$) (Toomre 1964).
This is expressed in terms of the Toomre parameter $Q \sim \sigma\Omega/\Sigma$
being smaller than unity.  If the inflow in the disc is driven by the 
instability, and in turn it provides the power necessary for stirring up 
$\sigma$, the disc tends to self-regulate itself at $Q \sim 1$ as follows.
When $Q$ drops well below unity, because of a high $\Sigma$ and/or a low 
$\sigma$, the disc becomes highly prerturbed, which increases the inflow rate.
This depletes $\Sigma$ and stirs up $\sigma$, pushing $Q$ to above 
unity. As a result, the fragmentation stops, the inflow slows down, 
so $\Sigma$ piles up by new accretion and $\sigma$ cools down, 
driving $Q$ to below unity, and so on.
The new element incorporated in our current analysis is treating the 
instability in a disc
made of two components, one dissipative and the other dissipationless,
including the continuous convertion of gas to stars.
The generalized instability analysis described in \se{instability}
refers to a two-component analog of the Toomre parameter, $Q_{\rm 2c}$, 
which is a function of the same $\Omega$, 
the surface densities of the two components, 
and their different velocity dispersions, 
$\sigma_{\rm stars} > \sigma_{\rm gas}$.
Since $\sigma_{\rm stars}$ cannot decrease and may actually increase in
time, the maintenance of $Q_{\rm 2c}\sim 1$ has to rely on $\sigma_{\rm gas}$ 
being low. As the gas mass fraction in the disc gradually declines, 
$\sigma_{\rm gas}$ has to get lower and lower.
However, when it becomes comparable to the thermal speed of sound, 
$c_{\rm s} \sim 10 \kms$, the pressure cannot keep decreasing as necessary,
and $Q$ is forced to be above unity. We refer to this as the end of the
violent instability phase, or ``stabilization", and attempt to find out
when it occurs for different choices of our model parameters.

\subsection{Mass Conservation}
\label{sec:massconserv}

Following \equ{mass}, the rate of change of gas mass in the disc is assumed to
be given by
\be
{\dot \Mdg} \simeq \Mgacc - \Mgev - {\dot M}_{\rm SFR} (1+\gamout) \, .
\label{eq:Mdd}
\ee
Here, the source term, $\Mgacc$, is the gas supply rate into the disc.
The sink term $\Mgev$ is the gas mass inflow rate in the disc towards the
central bulge, which we sometimes loosly refer to as ``migration".
The other sink term, ${\dot M}_{\rm SFR}$, includes the star formation rate
(SFR), and the outflow rate due to stellar feedback, 
which is assumed to be 
proportional to the SFR with $\gamout$ a parameter of order unity. 
Similarly, the rate of change of disc stellar mass is
\be
{\dot \Mds} \simeq \Msacc - \Msev + {\dot M}_{\rm SFR}\, ,
\label{eq:Msd}
\ee 
where $\Msacc$ represents the stellar fraction in the accretion rate into the 
disc, $\Msev$ is the stellar mass inflow rate within the disc,
${\dot M}_{\rm SFR}$ is the same as in \equ{Mdd} but with an opposite sign, 
and the stars are assumed not to be affected by outflows.
The rate of change of the total disc mass is the sum of (\ref{eq:Mdd}) and 
(\ref{eq:Msd}).
The source terms of accretion and the SFR are estimated as described below. 
Then, by appealing to energy conservation, gas-turbulence dissipation, 
and self-regulated instability, we will determine the inflow rates within the 
disc, and will be in a position to integrate \equ{Mdd} and \equ{Msd}. 

For completeness, although we do not explicitly integrate it with time, 
we note that the growth of the central bulge 
is described by
\be
{\dot M}_{\rm blg} = {\dot M}_{\rm blg,mer} + {\dot M}_{\rm bar, inflow} \, ,
\label{eq:Mblg}
\ee
where ${\dot M}_{\rm blg,mer}={\dot M}_{\rm bar} - \Mgacc -\Msacc$ 
is the fraction of the total cosmological input 
of baryons that come in as big mergers that build the bulge without joining the
disc, and ${\dot M}_{\rm bar, inflow} = \Mgev + \Msev$ is the total
baryon inflow rate within the disc (\se{cosmoaccr}).
We define the bulge-to-disk ratio as
\be
{\rm B/D} \equiv \frac{M_{\rm blg}}{\Mds+\Mdg} \, .
\ee

Following DSC09, we appeal to the convenient parameter $\deld$, 
the fraction 
of mass in the disc component within the characteristic radius of the disc, 
$\Rd$,
\be
\deld \equiv \frac{\Mds+\Mdg}{\Mt} \, .
\label{eq:f1}
\ee
Here the total mass $\Mt$ within $\Rd$ includes the contributions of gas and 
stars in the disc,
the stellar bulge, and the dark matter within $\Rd$.
The maximum possible value of $\delta$ is $\beta$, the fraction 
of baryons including disc and bulge within the disc radius,
\be
\deld \leq \beta \equiv \frac{\Mb}{\Mt} \, .
\label{eq:beta2}
\ee 
The ratio of disc to total baryonic mass is then $\Md/\Mb = \beta^{-1}\deld$, 
so that a bulge-less disc corresponds to $\deld = \beta$,
and a disc-less bulge is $\deld =0$.
The bulge-to-disc ratio is ${\rm B/D} =(\beta-\deld)/\deld$.
For gas and stars separately, we have
\be 
\delg \equiv \frac{\Mdg}{\Mt}  
\qquad {\rm and} \qquad \dels \equiv \frac{\Mds}{\Mt} \, ,
\label{eq:f2}
\ee
with $\deld = \delg + \dels$. 
In order to estimate the value of $\beta$, following DSC09,
we crudely treat the halo as an isothermal sphere with a mass profile 
$M(r) \propto r$, 
so we can write $\Mt \simeq \lambda M_{\rm vir} + M_{\rm bar}$, where 
$\lambda$ is the halo spin parameter, $M_{\rm vir}$ is the halo virial mass, 
and $M_{\rm bar}$ is the baryonic mass. This leads to 
$\beta \simeq f_{\rm bar}/(f_{\rm bar} + \lambda)$,
where $f_{\rm bar}$ is the baryonic mass fraction within the virial radius 
(which could be lower than the universal fraction due to mass loss in 
outflows).  Throughout the paper we assume 
$\lambda = 0.05$, $f_{\rm bar} = 0.075$, and thus $\beta = 0.6$.

\subsubsection{Cosmological Gas Supply}
\label{sec:cosmoaccr}

Both analytic estimates and cosmological simulations predict that the cosmological baryonic input 
funnels into high-redshift galaxies through cold 
streams that follow the filaments of the cosmic web and include merging 
galaxies and a smoother component
(e.g., Birnboim \& Dekel 2003; Keres et al. 2005; Dekel \& Birnboim 2006; 
Ocvirk et al. 2008; Dekel et al. 2009).
The average baryon input
rate is well approximated by the universal baryon fraction times the 
total cosmological input rate into galactic haloes 
(Dekel et al. 2009).
We estimate the corresponding timescale for accretion into a halo of mass 
$M_{\rm vir}$ at redshift $z$ by 
 \be
t_{\rm acc} \equiv \frac{M_{\rm vir}}{{\dot M}_{\rm vir}} 
\simeq 2.1 \, (1+z)_3^{-2.4}\, M_{12}^{-0.14}\, \Gyr \, ,
\label{eq:Mdot}
\ee
where $(1+z)_3\equiv (1+z)/3$ and $M_{12}\equiv M_{\rm vir}/10^{12}\msun$.
This approximation for the average specific accretion rate has been derived 
by fine tuning an analytic prediction based on the EPS approximation
(Neistein et al. 2006), and it has been shown to fit well the halo growth 
rate measured in the Millennium cosmological N-body simulation 
(Neistein et al. 2008; Genel et al. 2008; Fakhouri et al. 2010),
except that the numerical coefficient and the powers in \equ{Mdot}
were slightly adjusted for the WMAP7 cosmological parameters used here.
\Equ{Mdot} is accurate to better than 5\% in the redshift range $0.2 < z <5$,
and is an underestimate of $\approx 10\%$ at $z=0$ and $z=10$. 

The analytic EPS prediction in the high-redshit Einstein-deSitter 
cosmological regime is actually $\dot{M} \propto (1+z)^{5/2}$. 
This is a good approximation at $z>1$, where the redshit
is related to the Hubble time in Gyr as $(1+z) \simeq 6.6 t^{-2/3}$.
For the purpose of
a toy model that will turn out useful for 
back-of-the-envelope estimates, we tentatively 
ignore the weak mass dependence of \equ{Mdot}, 
and keep the same specific accretion rate for $M_{12}=0.56$ at
 $(1+z)=3$. 
By integrating \equ{Mdot}, such a halo ends up as a Milky-Way halo 
with $M_{12}=2$ at $z=0$,
which is comparable to the final mass of the halo used in our fiducial 
model below.
The toy-model version of \equ{Mdot} becomes 
\be
\frac{\dot{M}}{M} \simeq A\,(1+z)^{5/2}\, , \quad A \simeq 0.028 \Gyr^{-1}\, . 
\label{eq:Mdot_toy}
\ee
This implies a simple expression for the halo and galaxy mass growth, 
\be
M \propto e^{-\alpha\,z} \, ,
\label{eq:Mexp}
\ee
with $\alpha \simeq 25.4\, A \simeq 0.72$. 
A similar exponential growth has been found using 
cosmological N-body simulations (Wechsler et al. 2002). 

In our model, we assume that a fraction $\gamacc$
of the average baryonic mass input rate actually joins the disc as gas.
This is a multiple of two factors, 
(1) the fraction of the baryonic input in smooth accretion 
including small mergers that join the rotating disc, as opposed to more massive
mergers that build the bulge, and (2) the fraction of gas in this smoother
component, as opposed to stars that come in with small merging galaxies
and also join the disc.
We thus write for the specific gas accrertion rate into the disc
\be
\label{eq:gamacc}
\left.
\frac{{\dot M}_{\rm bar, acc}}{M_{\rm bar}}
\right|_{R_{\rm disc}}
\simeq
\gamacc \,
t^{-1}_{\rm acc}
\equiv
t^{-1}_{\rm disc, acc} \, .
\ee 
We test the effects of varying $\gamacc$ about a fiducial 
value of $\gamacc=0.7$. This is based on the estimate of 
D09 from hydro cosmological simulations that the average fraction
of incoming mass in clumps that lead to mergers with a mass ratio larger 
than 1:10 is about 30\%. 

We have made so far three simplifying assumptions that are worth mentioning.
First, 
we limit the current analysis to the simple case of accretion at the average 
cosmological rate. 
In reality, the accretion rate is varying, both among different galaxies
and along the history of each galaxy. The effects of these
variations in the accretion rate will be studied in a follow-up paper.
Second, 
the accretion rate onto galaxies may be sensitive to uncertain feedback 
processes, perhaps more important at lower redshifts (e.g., van de Voort et al.
2010).  A detailed modeling of this is beyond the scope of this paper.
Third, 
a fraction of the accreting mass into the disc is expected to be already in 
stars, formed earlier in small merging galaxies 
(e.g., D09; Ceverino, Dekel \& Bournaud 2010). 
These stars could partly contribute to a ``cold" stellar component that would 
participate in the gravitational disc instability and partly to a ``hot" 
stellar component, equivalent to the bulge
in terms of its effect on the disc instability.
In our current application we do not explicitly account for the stars that 
accrete onto the cold disc, and absorb the corresponding uncertainty in the 
value of $\gamacc$, assumed to represent the fraction of mass 
that accretes onto the disc, all in gas.

\subsubsection{Star Formation \& Outflows}
\label{sec:sfr_outflows}

As summarized in Krumholz, Dekel \& McKee (2011) and references therein, 
stemming from the Kennicutt-Schnidt empirical SFR law (e.g., Kennicutt 1998),
the star formation rate can be best modeled by a universal volumetric local 
star formation law.
When expressed in terms of surface densities it has the form
\be
{\dot \Sigma}_{\rm SFR} 
\equiv \frac{\Sigg}{\tau_{\rm SFR}}\,  
= \epssfr \frac{\Sigg}{t_{\rm ff}}\, ,
\label{eq:sfr}
\ee
where ${\dot \Sigma}_{\rm SFR}$ is the SFR surface density,
$\Sigg$ is the mass surface density of molecular gas,
and $t_{\rm ff} = [3\pi/(32 G \rho)]^{1/2}$ is the local 
free-fall time in the star forming region, derived from the local volumetric 
mass density $\rho$.  The dimensionless parameter $\epssfr$ is the 
SFR efficiency, i.e., the fraction of gas 
that is transformed into stars per free-fall time, and has been argued based 
on theory and observations to be constant in all star forming regions, 
at the level of 1 to a few percent.
In high redshift discs that are self-regulated at a Toomre instability with 
$Q \sim 1$, the free-fall time in the giant clumps where stars form
are comparable to the global disc crossing time $t_{\rm dyn}$,  
the inverse of the disc angular velocity $\Omega=V_{\rm circ}/\Rd$, 
which is comparable to its vertical crossing time 
(Krumholz, Dekel \& McKee 2011). 

In a variant of \equ{sfr}, Krumholz et al. (2009, hereafter K09) have argued 
that the physics of star formation 
within a molecular cloud, and in particular the variation in $t_{\rm ff}$,
can be encapsulated in the following star formation law
\be\label{eq:K09}
{\dot \Sigma}_{\rm SFR} = \frac{\Sigg}{2.6\Gyr}
\times 
\left\{
\begin{array}{ll}
\Sig85^{-0.33} \quad {\rm if} \quad \Sig85 < 1 \nonumber
\\
\Sig85^{+0.33} \quad {\rm if} \quad \Sig85 > 1 \, ,
\end{array}
\right.
\ee
where $\Sig85 = \Sigg/(85 \Msun\pc^{-2})$.
The origin of these two regimes is that clumps in discs with $\Sig85 > 1$ 
are in the regime where they have the characteristic Toomre mass, 
while in galaxies of lower surface density the clumps collapse and fragemnt 
until they reach the critical surface density of 
$85 \Msun\pc^{-2}$ that characterize low-redshift molecular clouds. 
The galactic discs at high redshift are practically always in the regime 
where $\Sig85 > 1$ while at low redshift they enter the other regime. 
In relatively massive, metal rich discs, the gas in the disc can be assumed 
to be all molecular, while in less massive discs, especially at high redshift, 
the molecular gas fraction becomes smaller than unity and should be 
considered when deriving the molecular surface density from
the total gas surface density (Krumholz \& Dekel 2011).

Outflows due to stellar feedback are incorporated in \equ{Mdd} through the 
term 
\be\label{eq:gamout}
{\dot M}_{\rm out} = \gamout\,{\dot M}_{\rm SFR}
\ee
with $\gamma_{\rm out}$
assumed to be of order unity and constant with time (see Table~2). 
This parameterization is based on theoretical and
observational results  
(Bouch{\'e} et al. 2006; 2007; 2009; Martin \& Bouch{\'e} 2009). 
While this takes into account the important role of feedback
in removing gas from the disc, 
we assume that the contribution of feedback to stirring up gas 
turbulence on galactic scales is minor 
(Joung et al. 2009; Ostriker \& Shetty 2011; but see also Hopkins et al. 2011).

\subsection{Energy Conservation}
\label{sec:energy}

The gravitational instability in the disc
is associated with torques that drive an outward angular-momentum flux 
and a corresponding mass inflow towards the disc centre 
(Gammie 2001; DSC09; Krumholz \& Burkert 2010; Bournaud et al. 2011). 
This inflow involves both the gas and the stars, in the form of clump 
migration as well as 
inflow of the smoother material in the disc outside the clumps. 
We assume that the gravitational energy released by this mass inflow down 
the potential well,
at a rate ${\dot E}_{\rm gas, inflow}  + {\dot E}_{\rm star, inflow}$, 
is deposited in the velocity dispersions of the two disc components.
This is a {\it ``gravitational heating"} effect
(Birnboim \& Dekel 2008; Khochfar \& Ostriker 2008).
While the stellar component does not dissipate the acquired energy,
the internal energy of the gas component dissipates on the turbulence 
dissipation timescale at a rate ${\dot E}_{\rm dis}$.
This is summarized in an energy conservation equation for the disc internal 
energy,
\be
{\dot E}_{\rm disc, int} =
{\dot E}_{\rm gas, inflow}+{\dot E}_{\rm star, inflow}-{\dot E}_{\rm dis} \, .
\ee
In order to solve our system of equations, we need to use an additional 
physical constraint. 
Here we make the strong assumption, that the inflow velocites of the gas and 
stars are the same, and that the energy gain is split between the two 
components in proportion to their masses. An example of stellar inflow is the 
clumps migration, but the inter-clump stars are also
flowing in. Such a behavior is indicated in zoom-in cosmological simulations 
(Cacciato et al., in preparation), but for now it should be considered a 
working assumption, to be fine-tuned later.
This assumption allows us to split the energy equation into two, 
\be
{\dot E}_{\rm gas, int} = 
{\dot E}_{\rm gas, inflow} -  {\dot E}_{\rm dis} - {\dot E}_{\rm SFR} 
\ee
\be
{\dot E}_{\rm star, int} = 
{\dot E}_{\rm star, inflow} + {\dot E}_{\rm SFR} \, ,
\ee
where ${\dot E}_{\rm gas, int}$ and ${\dot E}_{\rm star, int}$ represent 
the rate of change of internal energy for gas and stars, respectively, 
and ${\dot E}_{\rm SFR}$ refers to the energy
transfer between the two components as gas turns into stars. 

The gravitational potential difference between the outer disc edge and the 
disc centre is of order $V^2_{\rm circ}$. 
We then rewrite the energy equations as 
\be
\frac{3}{2}\frac{{\rm d} (M_{\rm gas} \sigma^2_{\rm gas})}{{\rm d}t} 
\approx V^2_{\rm circ} {\dot M}_{\rm gas, inflow}
- \frac{3}{2}M_{\rm gas} \sigma^2_{\rm gas}
(t_{\rm dis}^{-1} - \tau_{\rm SFR}^{-1})
\label{eq:energyg}
\ee
\be
\frac{3}{2}\frac{{\rm d} (M_{\rm star} \sigma^2_{\rm star})}{{\rm d}t} 
\approx V_{\rm circ}^2 {\dot M}_{\rm star, inflow}
+ \frac{3}{2} \frac{M_{\rm gas} \, \sigma^2_{\rm gas}}{\tau_{\rm SFR}} \, .
\label{eq:energys}
\ee   
The internal velocity is associated with a one-dimensional gas velocity 
dispersion $\sigg$. 
It consists of a contribution from turbulence, $\sigma_{\rm turb}$, 
and from thermal energy, through the speed of sound $c_{\rm s}$,
$\sigg^2 = c^2_{\rm s} + \sigma^2_{\rm turb}$. 
We assume that $c_{\rm s} \simeq 10 \kms$, corresponding to gas cooled by 
atomic cooling to $10^4$K.
  
The turbulence dissipation timescale in \equ{energyg} is parametrized
as proportional to the disc vertical crossing 
time, which for $Q \sim 1$ is comparable to the crossing time in the disc 
plane, namely
\be
t_{\rm dis} = \gamdis t_{\rm dyn}\, .
\label{eq:gamdis}
\ee
The value of $\gamdis$
is expected to be about unity if the turbulence lengthscale $\ell_{\rm turb}$
is comparable to the disc scale height $h_{\rm disc}$ or smaller, 
while it could be somewhat larger, 
$\gamdis \simeq \ell_{\rm turb}/h_{\rm disc}$,  
if there is turbulence on scales larger than $h_{\rm disc}$
We adopt $\gamdis=3$ as our fiducial value, and study the effect of 
vaying this parameter between 1 and 10 in \S~4.1. 

Three further comments are in place regarding the energy equations.
First,
since the star formation timescale is two orders of magnitude larger than
the dynamical timescale, the dissipation term dominates the right hand side of
eq.~(\ref{eq:energyg}). 
Thus, as long as the internal energy of the gas is varying slowely, 
\equ{energyg} is approximated by \equ{approxenergy}, 
where the gas inflow rate is determined by the dissipation rate, 
with only a minor correction due to the presence of the stellar component.
Second, when the turbulence becomes very weak with a small $\sigma_{\rm turb}$,
the dissipation timescale becomes much longer than the dynamical time,
and the inflow rate slows down accordingly. In our simple model we do not 
explicitly incorporate this effect, and instead consider the time at which 
$\sigg$ becomes as small as 
$c_{\rm s}$ as the time of ``stabilization", which we term $z_{\rm stab}$.
Third, in \equ{energys}, the change in the stellar internal energy is assumed 
to consist of two terms.
The second term represents the addition of newly born stars with a velocity 
dispersion that equals the instantaneous gas velocity dispersion, while the 
first term corresponds to the gravitational ``heating" of the stars by the 
stellar mass inflow in the disc. 
It is worth recalling that the stellar heating rate is determined by the 
assumption that the disc stars flow in with the same velocity as the gas.

\subsection{Two-Component Gravitational Instability}
\label{sec:instability}
According to the standard {\it one-component\,} Toomre instability analysis 
(Toomre 1964), a thin 
rotating disc of gas or stars becomes unstable to axisymmetric modes
once the 
attraction due to self-gravity,
represented by the surface density $\Sigma$,
overcomes both the centrifugal force due to rotation
and the pressure force associated with the radial velocity dispersion $\sigma$.
For gas, there is an additional contribution from thermal pressure, but 
the high-redshift discs under investigation here are 
in a regime where the velocity dispersion associated with turbulence is 
larger than the speed of sound that characterizes the thermal pressure.
The instability is expressed in terms of the condition that the Toomre 
parameter $Q$ is smaller than 
a critical value $Q_{\rm crit}$ of a value about unity (Toomre 1964),
\be
Q = \frac{\sigma \kappa}{\pi G \Sigma} < Q_{\rm crit}  \simeq 1 \, .
\label{eq:Q1}
\ee
Here $\kappa$ is the epicyclic frequency, related to the angular circular 
velocity $\Omega(r)$ by $ \kappa^2 = r\, {d\Omega^2}/{dr} +4\Omega^2$.
For a power-law rotation curve $V_{\rm circ} \propto r^\nu$,
this becomes $\kappa^2 = 2(1+\nu)\Omega^2$.  We adopt here a flat rotation 
curve, as seen in cosmological simulations, namely $\nu=0$.
We adopt an effective value $\Omega = V_{\rm circ}/R_{\rm disc}$,
where the circular velocity is approximated by the virial velocity, 
and the disc radius is a constant fraction of the virial 
radius, $R_{\rm disc} = \lambda \, R_{\rm vir}$, with $\lambda = 0.05$ 
representing the halo spin parameter. Thus, the time evolution of $\kappa$ 
depends only on the cosmological evolution of the halo virial quantities (see \S~3).
For a thin disc, gaseous or stellar, $Q_{\rm crit} \simeq 1$.
For a thick disc, the Toomre analysis is valid as long as the length scale of 
the perturbation is larger than the disc thickness and smaller than the disc 
radius, with $Q_{\rm crit} \approx 0.67$ (Goldreich \& Lynden-Bell 1965).

When the disc is made of two components, gas and stars, each with different 
$\Sigma$ and $\sigma$, one can address the axisymmetric instability via a 
two-dimensional Toomre-like parameter $Q_{\rm 2c}$
which can be expresses in terms of $Q_{\rm star}$ and $Q_{\rm gas}$, 
each defined by \equ{Q1} with the $\Sigma$ and $\sigma$ 
of the corresponding component
(Jog \& Solomon 1984; Romeo 1994; Wang \& Silk 1994; Rafikov 2001).
The difference in the expresion for $Q_{\rm 2c}$ due to the dissipative nature 
of the gas is small when we focus attention to the most unstable scale 
(Rafikov 2001), and we ignore it here.
Romeo \& Wiegert (2011) proposed the convenient expresion that we use here,
\be
Q^{-1}_{\rm 2c} =
\left\{
\begin{array}{ll}
& \!\!\!\!\!\!\!\!\!
W \, {\cal Q}^{-1}_{\rm star} 
+ {\cal Q}^{-1}_{\rm gas} \quad {\rm if} \quad {\cal Q}_{\rm star} > 
{\cal Q}_{\rm gas}\\
 & \\
& \!\!\!\!\!\!\!\!\!
{\cal Q}^{-1}_{\rm star} 
+ W \, {\cal Q}^{-1}_{\rm gas} \quad {\rm if} \quad {\cal Q}_{\rm star} <
{\cal Q}_{\rm gas} \, ,
\end{array} \right.
\label{eq:Q2}
\ee
where 
\be
W = \frac{2 \sigs \sigg}{\sigs^2 +\sigg^2} \, .
\label{eq:W}
\ee 
Two-component instability is characterized by $Q_{\rm 2c} \leq 1$. 
The parameter $Q_{\rm 2c}$ can be thought of as a combination of $\Qs$ and 
$\Qg$, weighted by a function of the ratio of the velocity dispersions of gas 
and stars.

In order to account for a disc thickness, associated with a vertical velocity 
dispersion $\sigma_{i,z}$ for the $i$'s component (gas or stars), 
the quantities that enter \equ{Q2} are ${\cal Q}_i = T_i Q_i$, 
with the approximation
$T_{i} \simeq 0.8+0.7(\sigma_{i, z}/\sigma_{i})$ valid for 
$0.5<\sigma_{i,z} /\sigma_{i}<1$ 
(Romeo \& Wiegert 2011; Romeo 1994). 
Setting $T_{i}$ to unity corresponds to a thin disc in that component.
Here we adopt disc thicknesses that are determined by the assumption of 
isotropic velocity dispersion for each component, $\sigma_{i, z} = \sigma_{i}$.
As a sanity check, note that under this assumption ${\cal Q}_{\rm i} = 1$ 
corresponds to $Q_{\rm i} \approx 0.67$, in agreement with the original result 
for a one-component thick disc (Goldreich \& Lynden-Bell 1965).

Note that the two-component system can be unstable, $Q_{\rm 2c} \leq 1$,
even when each of the components has $Q_i > 1$.
For instance, in the thin-disc approximation, with the gas and stars having 
the same $\sigma_i$ and the same $\Sigma_i$, $Q_{\rm 2c} = 1$ corresssponds 
to $Q_i=2$ for each of the components.

\section{Model Predictions}
\begin{figure}
\centerline{\psfig{figure=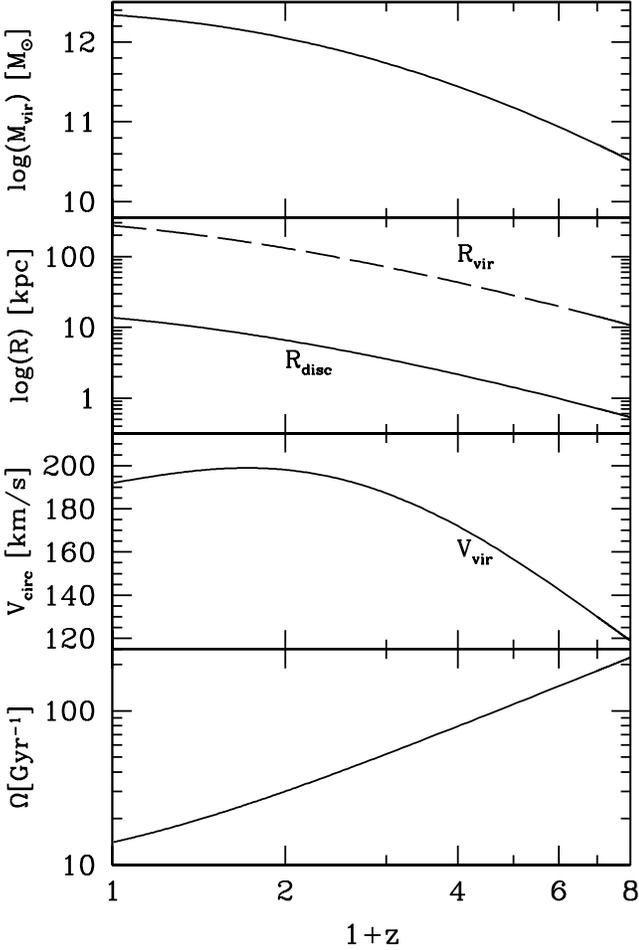,width=\hsize}}
\caption{
Cosmological evolution of the global quantities in our fiducial model galaxy. 
Shown from top to bottom are (a) the halo virial mass $M_{\rm vir}$, 
(b) the virial and disc radius, $R_{\rm vir}$ and 
$R_{\rm disc}=\lambda R_{\rm vir}$, with $\lambda=0.05$,
(c) the circualr velocity $V_{\rm circ}$, assumed equal to the virial velocity 
$V_{\rm vir}$, and (d) the angular velocity 
$\Omega = V_{\rm circ}/R_{\rm disc}$, which at high redshift is inversely
proportional to the cosmological time.
}
\label{fig:fig1}
\end{figure}

In the current paper we focus on a halo that grows by the average 
cosmological growth rate, \equ{Mdot},
into a halo comparable to the Milky Way halo with $M_{12}=2$ at $z=0$.
The virial relations between the halo virial
mass, velocity and radius are
\be
V^2_{\rm vir} = \frac{G M_{\rm vir}}{R_{\rm vir}}
\, ,
\quad
\frac{3 M_{\rm vir}}{4 \pi R^3_{\rm vir}}  = \Delta {\bar \rho}\, ,
\ee
where ${\bar \rho}$ is the average mass
density of the universe at that time.
In the $\Lambda$CDM cosmology, they can be expressed as
\be
M_{12} \approx 0.6 V^3_{100} A^{3/2} \approx 34.2 R_1^3 A^{-3} \, ,
\ee
where $V_{100} \equiv V_{\rm vir}/100$ km/s, 
$R_1 \equiv R_{\rm vir}/1$ Mpc, and $A$ is the modified expansion factor,
\be
A \equiv a \left[\frac{\Delta}{200} \frac{\Omega_{\rm m}}{0.3} 
\left(\frac{h}{0.7}\right)^2\right]^{-1/3} \, ,
\ee
with $a = 1/(1+z)$. 
The cosmological time evolution of 
the density parameter $\Omega_{\rm m}$ 
and the Hubble constant, $h=H_0/100 \kms\Mpc^{-1}$, 
is given by
\be
\Omega_{\rm m}(a) 
= \frac{\Omega_{\rm m} a^{-3}}{\Omega_{\Lambda} + \Omega_{\rm m}a^{-3}} ,
\quad
H(a) = H_0 \, (\Omega_{\Lambda} + \Omega_{\rm m} a^{-3})^{1/2} ,
\ee
and we use 
the approximation proposed by Bryan \& Norman (1998) 
for the cosmological time dependence of $\Delta$:
\be
\Delta = 18 \pi^2 - 82 \, \Omega_{\Lambda} -39 \,\Omega_{\Lambda}^2 \, .
\ee
The halo mass growth is shown in the top panel of \fig{fig1}.
It roughly follows the approximation \equ{Mexp}, reaching $M_{12} \simeq 0.5$
at $z=2$ and $M_{12} \simeq 1$ at $z=1$.
 
The second panel of \fig{fig1} shows the corresponding $R_{\rm vir}$
and $R_{\rm disc}=\lambda\,R_{\rm vir}$ with $\lambda=0.05$. 
The third panel of \fig{fig1} shows the evolution of the corresponding 
disc circular velocity, 
which we assume equals the virial velocity. 
This velocity is growing with time to a flat maximum at $z\simeq 1.1$, 
followed by a gradual decline towards $z=0$.
This can be reproduced with the mass growing as in the approximation 
\equ{Mexp},  
\be
\label{eq:Vvir}
V_{\rm vir} \propto (1+z)^{1/2} \, \exp(-\alpha z/3)\, .
\ee
This indeed reaches a maximum at $z_{\rm max}\simeq 3/(2\alpha)-1 \simeq 1.1$.
We will see below that this governs the evolution of $\sigma$ 
in the one-component case, and that it also has an important effect 
on the evolution of the velocity dispersions in the two-component case, 
partly through the dependence of the gravitational heating on the potential 
well expressed by $V_{\rm circ}$.
 
The bottom panel of Fig.~1 shows the angular velocity $\Omega$, 
which enters in the 
instability $Q$ parameter. 
Recall that $\Omega = t^{-1}_{\rm dyn}$. In the Einstein deSitter phase where $\Delta_{\rm vir}$ is constant, and under the assumption of a constant spin parameter $\lambda$, $t_{\rm dyn}$ is proportional to the 
cosmological time\footnote{
For the standard 
$\Lambda$CDM cosmology, 
in its Einstein-de Sitter
regime, the virial radius and velocity are related to the virial mass as 
$R_{\rm vir} \sim 308$ kpc
$(1+z)^{-1} \, M^{1/3}_{12}$ and 
$V_{\rm vir} \sim 118 $ km/s $(1+z)^{1/2} \, M^{1/3}_{12}$. 
Thus $V_{200} \sim R_{100} (1+z)_3^{3/2}$, 
where the quantities are in units of 200 km/s and 100 kpc.
}.

\subsection{A one-component disc}
\label{sec:onecomp}

\begin{figure}
\centerline{\psfig{figure=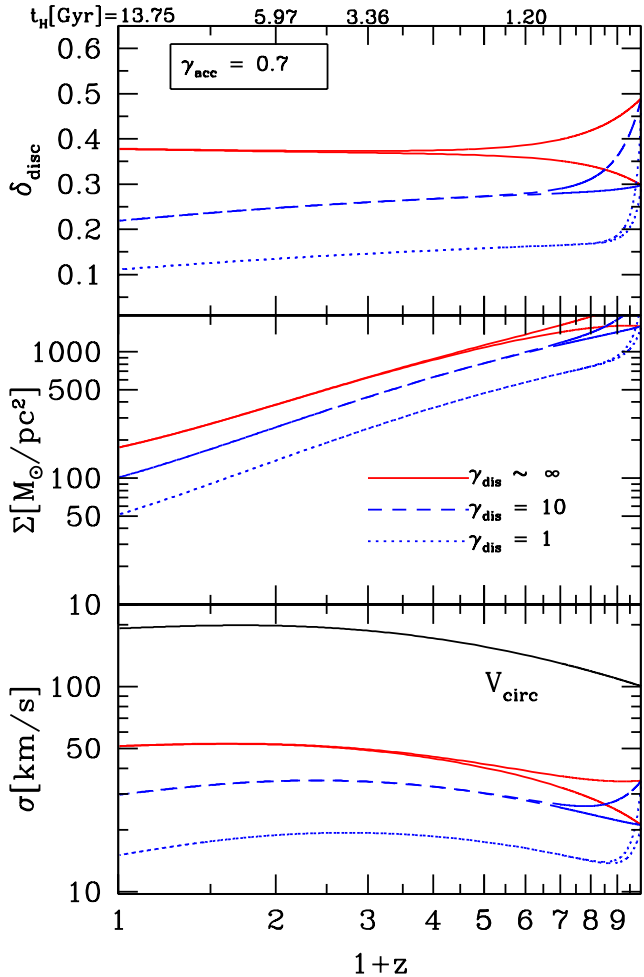,width=\hsize}}
\caption{
Evolution of a one-component disc and the role of dissipation.
Red solid lines refer to non-dissipative stellar discs, and blue dashed 
and dotted lines refer to gaseous discs with $\gamdis =10$ 
and $1$, respectively.
{\it Top:}
Mass fraction in the disc within $R_{\rm disc}$, $\delta_{\rm disc} \simeq \sqrt{2} \sigma/V_{\rm circ}$.
{\it Middle:}
Disc surface density $\Sigma$.
{\it Bottom:}
Velocity dispersion $\sigma$ and $V_{\rm circ}$ for comparison.
Lower $\gamdis$, corresponding to a higher dissipation rate,
requires a lower velocity dispersion for maintaining $Q=1$.
}
\label{fig:fig2}
\end{figure}

As a first application of our model, we investigate the evolution of 
instability in a one-component disc. Isolating $\sigma$ form \equ{Q1}, 
we can write for a disc with a flat rotation curve
\be
\label{eq:sigmagasonly}
\sigma 
= \frac{\pi G \Sigma_{\rm disc} Q_{\rm crit}}{\sqrt{2} \Omega} 
= \frac{1}{\sqrt{2}}  \deld V_{\rm circ} Q_{\rm crit}\, .
\ee
Inserting this expression for $\sigma$ in the mass and energy equations, 
we can straightforwardly integrate them.

Fig.~\ref{fig:fig2} shows the evolution of a one-component disc for different 
values of the dissipation parameter $\gamdis$ (\equ{gamdis}).
A choice of $\gamdis \sim \infty$ refers to a dissipationless, 
all stellar disc.  The finite values of $\gamdis$ in the range 10-1 
correspond to a pure gaseous disc with slow and rapid dissipation, 
$\gamdis=10,1$ respectively.
For each choice of $\gamdis$ we show the results for two different 
intitial conditions, which practically converge to a single solution by 
$z \sim 5$, in accordance with the exponential convergence described in 
\se{model}.

The top
panel of Fig.~\ref{fig:fig2} shows the time evolution of the disc 
mass fraction, $\deld$, which in the one-component analysis is 
$\sqrt{2} Q_{\rm crit} (\sigma/V_{\rm circ})$. 
We see that $\deld$ is roughly constant in time, or is varying 
rather slowly, indicating that the system evolves in a quasi-steady-state, 
where the disc and the bulge grow together, as emphasized in DCS09.
In the case of a stellar disc, the disc fraction is high, $\deld 
\approx 0.4$, close to the maximum possible value of $\beta = 0.6$, namely 
a bulge-to-disc ratio of 0.5. 
This reflects the fact that without dissipation the inflow rate in the disc 
is rather slow, \equ{approxenergy}.
For lower values of $\gamdis$, the disc fraction $\deld$ 
becomes lower, refelecting the higher depletion rate by inflow when the 
dissipation is faster.  With $\gamdis=10$, the steady-state is with 
$\deld \simeq 0.25$, namely B/D $\simeq 1.4$, and with 
$\gamdis=1$ it is $\deld \simeq 0.15$, 
corresponding to B/D $\sim 3$.

The middle panel of Fig.~\ref{fig:fig2} shows how $\Sigma_{\rm disc}$
is gradually decreasing with time. This reflects the assumptions
that the galaxy mass follows the halo mass growth,
that is close to exponential following \equ{Mexp},
and that the galaxy size is a constant fraction of the virial radius.
Then, according to the approximate behavior of \equ{Mexp} and in the 
Einstein-deSitter regime with a constant $\lambda$, 
the three-dimensional density is proportional to the virial density, 
namely it is independent of mass and follows the cosmological density 
evolution. We thus have: 
\be
\Sigma \propto \frac{M}{R^2} 
\propto \deld \, (1+z)^2\, \exp(-\alpha z/3)\, .
\label{eq:Sigma}
\ee
For a slowly varying $\deld$,
as seen in the top panel of \fig{fig2},
and for $\alpha \simeq 0.72$ as in \equ{Mexp}, 
the surface density of \equ{Sigma}
is indeed decreasing with time at all redshifts $z<7.4$, 
namely throughout most of the relevant redshift range.

\begin{figure}
\centerline{\psfig{figure=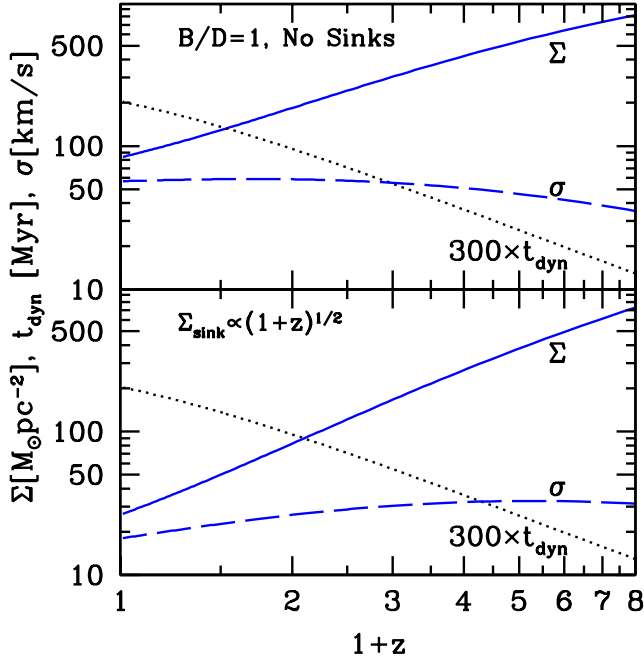,width=\hsize}}
\caption{
An illustration of the way the velocity dispersion compensates for the
rate of change of $\Sigma/\Omega$ to keep $Q \sim 1$. 
{\it Top:} a case where the disc grows by cosmological accretion
but all sink terms are turned off, showing $\sigma \propto \Sigma/\Omega
\simeq\const$.
{\it Bottom:} the same but with an arbitrary sink term added, 
showing a decline of $\sigma$ in the regime where $\Sigma/\Omega$ declines,
similar to \fig{fig2}.
}
\label{fig:fig1b}
\end{figure}

The lower panel of Fig.~\ref{fig:fig2} shows the evolution of the
velocity dispersion in the one-component models, with $V_{\rm circ}$ shown as
a reference. 
Based on eq.~(\ref{eq:sigmagasonly}), 
$\sigma$ is determined by $\Sigma/\Omega$,
or equivalently by $\deld\, V_{\rm circ}$. 
Since $\deld$ is rather constant, $\sigma$ is roughly proportional 
to $V_{\rm circ}$, which follows the evolution of the virial velocity
shown in \fig{fig1} and approximated in \equ{Vvir}.
This explains why $\sigma$ is rising to a flat maximum near $z \sim 1$
and is gradualy declining afterwards.
Thus, the evolution of $\sigma$ is driven by the cosmological halo growth rate
and the evolution of its virial velocity.
More precisely, the evolution of $\sigma$ is determined by the interplay between 
the rates of change of $\Sigma$ and $\Omega$ under the constraint that 
$Q \simeq 1$ (\equ{Q1}).
\Fig{fig1b} demonstrates this, and in particular the qualitative role played
by the sink terms in causing the decline of $\sigma$ at late times.
The top panel shows a simplified model in which mass accretes onto the disk
in the average cosmological rate,
but the sink terms are all turned off, $\delta_{\rm disc}=\const$,
and there is no dissipation. In this case, the evolution of $\Sigma$ and
$\Omega$ are dictated solely by the cosmological evolution of the virial
quantities, under the assumption of a cosntant spin parameter and 
$V_{\rm circ} = V_{\rm vir}$.
We note that in this case, in the redshift range $z=3-0$, 
that the decline rate of $\Sigma$ and $\Omega$ are quite similar, so $\sigma$
maintains a rather cosntant value. The toy model of \equ{Mexp},  
in the Eisntein-deSitter regime, indeed predicts a slow evolution 
$\sigma \propto \Sigma/\Omega \propto (1+z)^{1/2} e^{-0.24z}$.
The bottom panel of \fig{fig1b} refers to the same model but with
an artificial suppression term $(\propto (1+z)^{1/2})$. 
In this case, at late times, $\Sigma$ drops in time faster than $\Omega$,
so $\sigma$ has to decline in order to maintain $Q \sim 1$, 
as it does in \fig{fig2}.
\begin{table}
\label{tab:modpar2}
\caption{One-Component Models}
\begin{center}
\begin{tabular}{ccccc}
\hline
\hline
 & $\gamdis$ &  $\gamacc$ &
 $\tau_{\rm SFR}$& $\gamout$\\
\hline
Dissipation 
& $ \sim \infty$ & $0.7$ & --& 0 \\
& $10$ & $0.7$ & --& 0 \\
& $1$ & $0.7$ & -- & 0\\
\hline
\hline
\end{tabular}
\end{center}
\medskip
\begin{minipage}{\hsize}

\end{minipage}
\end{table}

For the models shown in \fig{fig2}, we see that self-regulated instability is
maintained till $z = 0$, in the sense that the model has
$\sigg > c_{\rm s} \simeq 10 \kms$ at all times, even 
for the case in which the dissipation rate
is as high as the disc crossing time. 
This result is consistent with the analysis of DSC09, 
who concluded that an unstable disc accreting at the average cosmological rate 
remains unstable throughout its cosmic history.

\begin{figure}
\centerline{\psfig{figure=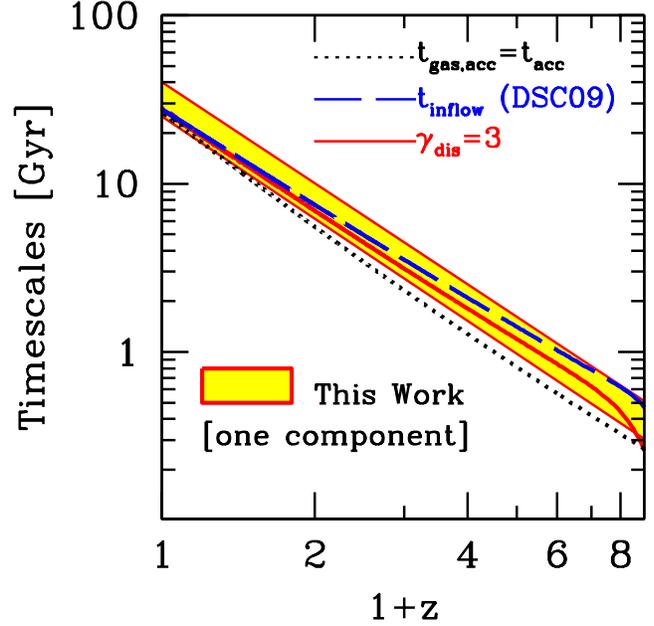,width=\hsize}}
\caption{
Evolution of relevant timescales, showing 
the cosmological accretion timescale (black dotted line),
the timescale for mass inflow in the disc as derived by DSC09 
from clump interactions and torques (blue dashed line),
the timescale for mass inflow in the disc according to our fiducial model
(red solid line), and its range corresponding to variations in $\gamdis$ 
in the range $(1,10)$ (yellow shaded area).
}
\label{fig:fig3}
\end{figure}

We note that DSC09 estimated the migration timescale via an explicit 
calculation of the effect of encounters between giant clump, and alternatively via torques in a viscous disc,
without considering the energy budget. 
In Fig.~\ref{fig:fig3} we show that our current model, based on energy considerations, without specifying the mechanism by
which this energy is transferred to velocity dispersion and to inflow in the 
disc, predicts
an inflow timescale similar to the estimate in DSC09. 
This agreement is non-trivial given the fundamental
difference between the two calculations. We also see that the two calculations
predict inflow timescales that are comparable to the accretion timescale.

%

\subsection{Disc of Gas and Stars: Fiducial Model}
\label{sec:twocomp}
\begin{table}
\label{tab:modpar2}
\caption{Two-Component Models}
\begin{center}
\begin{tabular}{ccccc}
\hline
\hline
 & $\gamdis$ &  $\gamacc$ &
 $\tau_{\rm SFR}$& $\gamout$\\
\hline
{\bf Fiducial}
&{\bf 3} &  {\bf 0.7} & {\bf K09} & {\bf 0}\\
\hline
Gas Dissipation
&$10$ &  $0.7$ & K09 & 0\\
&$1$ &  $0.7$ & K09 & 0\\
\hline
Gas Accretion 
&$3$ &   $1$ & K09 & 0\\
&$3$ &   $0.4$ & K09 & 0\\
\hline
Star Formation 
& $3$ & $0.7$ & $\tau_{\rm dyn}/0.01$ & 0\\
&$3$ & $0.7$ & $\tau_{\rm dyn}/0.05$ & 0\\
\hline
Outflows 
&$3$ &   $0.7$ & K09 & 1\\
&$3$ &   $0.7$ & K09 & 3\\
\hline
Extreme
&$1$ &   $0.4$ & K09 & 3\\
\hline
\hline
\end{tabular}
\end{center}
\medskip
\begin{minipage}{\hsize}
\end{minipage}
\end{table}

After recovering the persistance of instability in the one-component models,
we proceed to the two-component case. Here, the gas fraction in the disc
is gradually declining with time, which may lead to stabilization
once the disc becomes dominated by a hot stellar component.
In this section, we describe the results for a specific set of model 
parameters, which we refer to as the fiducial model. In \se{variations}
we will address the effects of varying the model parameters about the
fiducial values.

\begin{figure}
\centerline{\psfig{figure=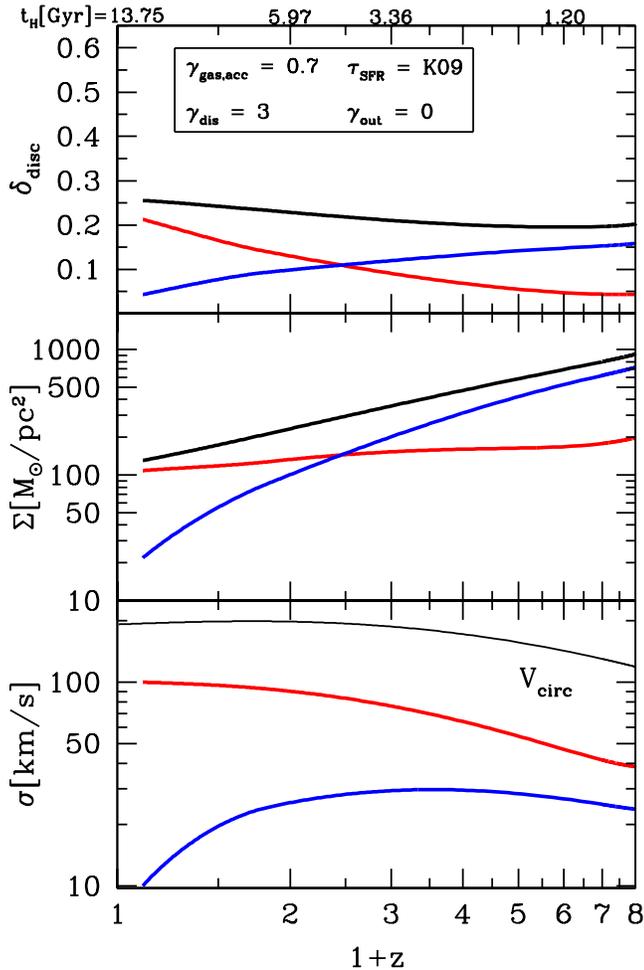,width=\hsize}}
\caption{
Evolution of the fiducial two-component model (Table~2).
Panels and curves are as in \fig{fig2}.
Shown are the quantities for the gas (blue), the stars (red), and their sum
(black).
The disc stabilizes at $z_{\rm stab} = 0.15$.
}
\label{fig:fig4}
\end{figure}

Our fiducial model, as specified in Table~2, consists of 
$\gamdis = 3$ (\equ{gamdis}), 
$\gamacc=0.7$ (\equ{gamacc}), 
the K09 star formation law (\equ{K09}),
and tentatively $\gamout=0$ (\equ{gamout}).
Fig.~\ref{fig:fig4} shows the time evolution of 
the same quantities shown in \fig{fig2}, namely $\deld$, 
$\Sigma$ and $\sigma$, but now, and in the following figures,
refering to the two components of gas and stars in blue and red respectively,
and to their sum in black.
We immediately note in the bottom panel that, unlike the always-unstable 
one-component models, the fiducial two-component model does stabilize 
shortly prior to $z=0$, at $z \simeq 0.15$, as marked by $\sigg$ 
droping below $c_{\rm s} = 10\kms$.
The drop is rather steep, starting at $z \sim 1$ after a pretty constant
$\sigg\simeq 30\kms$ till then.
It is driven by the disc turning from being gas dominated at early times
to star dominated at $z \simeq 1.5$, as seen in the top and middle panels,
with 
$\Sigs/\Sigg \simeq \delta_{\rm stars}/\delta_{\rm gas}
\simeq 5$ by $z=0.15$.
With the stars ``hot" at $\sigs \simeq 80-100\kms$ 
in the range $z \sim 1-0$, the gas has to cool faster than in the one-component
case in order to compensate for its decreasing fraction in the disc surface 
density and keep $Q_{\rm 2c}=1$.

We note several additional interesting features in the evolution of the 
fiducial model compared to the one-component models. 
First, the gas velocity dispersion is at the level of 
$\sigg \simeq 30\kms$ until it starts droping at $z \sim 1$.
At $z=2$ this corresponds to $V_{\rm circ}/\sigg \sim 6.6$,
reaching $V_{\rm circ}/\sigg \simeq 10$ at $z=0.15$.
The evolution of $\sigg$ in the two-component fiducial model is similar
to the evolution in the corresponding one-component gas model, except for the
drop at $z \leq 1$.
At $z=2$, $\sigs/\sigg \simeq 2.6$.
Second,
the total disc fraction within $R_{\rm disc}$ is rather constant in
time, evolving in a quasi steady state as in the one-component gas model, 
at a comparable level of $0.2-0.25$, i.e., B/D$\simeq 2-1.4$, but here
slowly rising instead of declining. 
The constancy of the bulge to disc ratio indicates that the massive bulge
is not the main reason for the stabilization 
Third, 
the stellar surface density in the fiducial model is almost constant 
throughout the disc evolution, contrary to its gradual declime in the
one-component stellar model.

\begin{figure}
\centerline{\psfig{figure=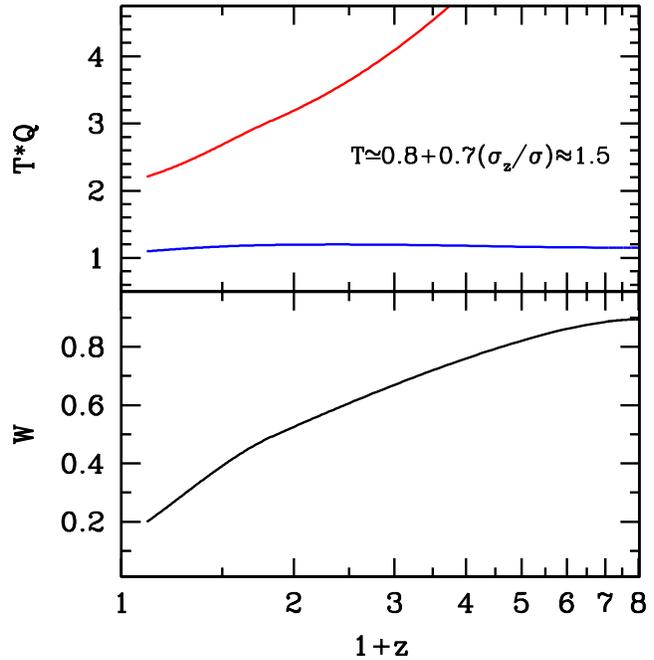,width=\hsize}}
\caption{
Evolution of the instability parameters in the fiducial two-component model. 
{\it Top:}
The $Q$ parameters, \equ{Q2}, for gas (blue) and for stars (red).
{\it Bottom:}
The weighting function $W$, \equ{W}.
With a lower value of $Q$, the gas is the main driver of
instability at all times.
}
\label{fig:fig4b}
\end{figure}

\Fig{fig4b} shows the evolution of the components of $Q_{\rm 2c}$,
\equ{Q2}, for the fiducial model.
It shows the values of ${\cal Q}=T Q$ for the gas and for the stars,
with $T \simeq 1.5$ under the assumption of isotropy adopted here,
and the value of $W$. 
We see that $Q_{\rm gas}$ is always smaller than $Q_{\rm stars}$.
This is because $\sigma_{\rm gas}$ is significantly smaller than 
$\sigma_{\rm stars}$, and especialy so when the surface density is dominated 
by the stars. For a similar reason $W$ evolves to well below unity.
As a result, $Q_{\rm gas}$ is dominant in determining $Q_{\rm 2c}$ 
(\equ{Q2}, top line), namely the instability is driven by the gas
even after the disc has become stellar dominated.

\section{Variations about the Fiducial Model}
\label{sec:variations}

We now explore the effects of varying the model parameters that govern the turbulence dissipation rate (\se{dissipation}), the fraction of the baryonic cosmological input that joins the disc as gas (\se{accretion}), 
the star formation rate (\se{sfr}), 
and the outflow rate (\se{outflow}).
Table~2 displays the values of the parameters used in the
different models.

\subsection{Gas Dissipation}
\label{sec:dissipation}

\begin{figure}
\centerline{\psfig{figure=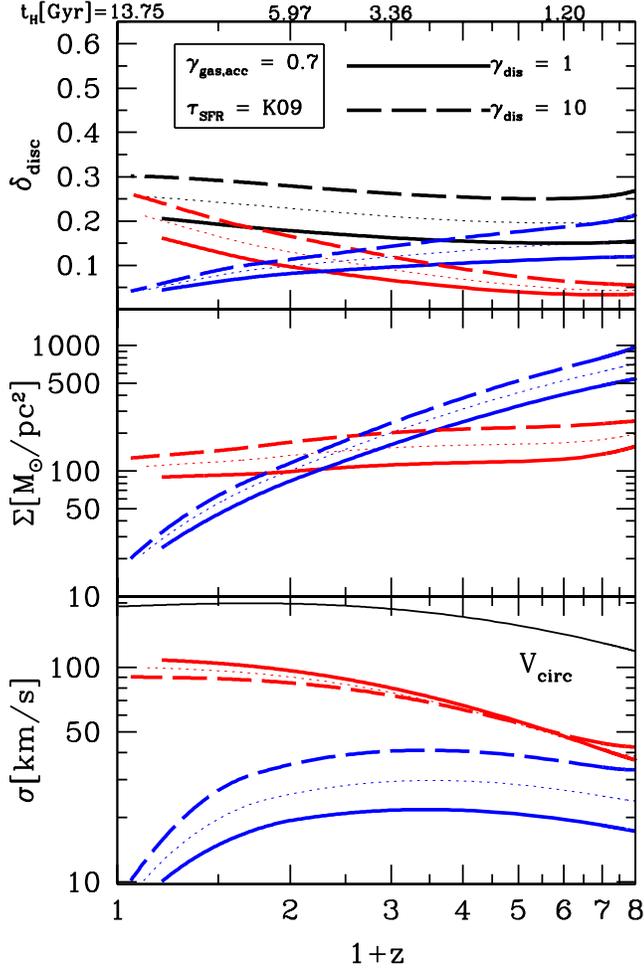,width=\hsize}}
\caption{
The effects of varying $\gamdis$ (\equ{gamdis}) in the range $1-10$
on the evolution of the two-component model.
Panels and curves are as in \fig{fig4}.
Dotted curves refer to the fiducial model of \fig{fig4} for comparison.
The stabilization epoch is only weakly affected,
but the gas dispersion velocity is significantly lower when the dissipation
rate is higher.
}
\label{fig:fig5}
\end{figure}

The energy considerations that lie at the basis of our model imply that
the mass inflow rate within the disc is determined by the gas turbulence 
dissipation rate (eq.~[\ref{eq:approxenergy}] and [\ref{eq:energyg}]) 
and that in turn the inflow down the potential gradient is the source 
of energy that 
adjusts the velocity dispersions of gas and stars
at the level that guarantees $Q_{\rm 2c} = 1$. 
In \fig{fig5}, we test the effect of varying the dissipation parameter 
$\gamdis$ (\equ{gamdis}) between the extreme values 1 and 10, 
bracketing the fiducial value $\gamdis=3$,
while keeping the other model parameters at their fiducial values.

We notice first that the dissipation rate has only a minor impact on 
the qualitative behavior, and in particular on the time of stabilization,
which shifts from $z_{\rm stab} \approx 0.1$ to $z_{\rm stab} \approx 0.3$
when $\gamdis$ is varied from 10 to 1, 
corresponding to a shift from a slow to a fast inflow rate.
A higher dissipation rate is responsible for a slightly earlier stabilization.
The reason for the small effect is that, by construction, the ratio 
$\Sigg/\Sigs$ is not very sensitive to the inflow
rate, that is driven by the dissipation rate, and this ratio is in fact 
slightly higher for faster dissipation. The transition to star dominance varies 
from $z \simeq 1.3$ to $1.6$ when $\gamdis$ is varied from 10 to 1.
Note that the surface densities of the two components are correlated not only
because of the shared inflow and the resultant depletion of the two 
components of the disc, but also because 
according to the SFR prescription $\Sigs$ is growing
in proportion to $\Sigg$. 
On the other hand, the value of $\gamdis$ has a non-negligible effect 
on the value of $\sigg$ prior to its drop at $z \leq 1$, which is 
$\simeq 40 \kms$ and $\simeq 20\kms$ for $\gamdis =10$ and 1, 
respectively.
The corresponding values of $V_{\rm circ}/\sigg$ at $z=2$ are $4.2$ and $7.4$
for $\gamdis=10-1$. 
While the drop of $\sigg$ may start a bit later for $\gamdis =10$,
the fact that it starts from a higher value makes it reach $10\kms$ at a 
slightly later time.

We also see, in the top panel, that a
higher dissipation rate, inducing a higher disc depletion rate,
makes the total disc fraction smaller, from 
$\deld\sim 0.25-0.3$ to $\sim 0.15-0.2$
when $\gamdis$ is varied from 10 to 1. These correspond to
B/D $\gsim 1$ and 2 respectively.

\subsection{Effective Accretion onto the disc}
\label{sec:accretion}

\begin{figure}
\centerline{\psfig{figure=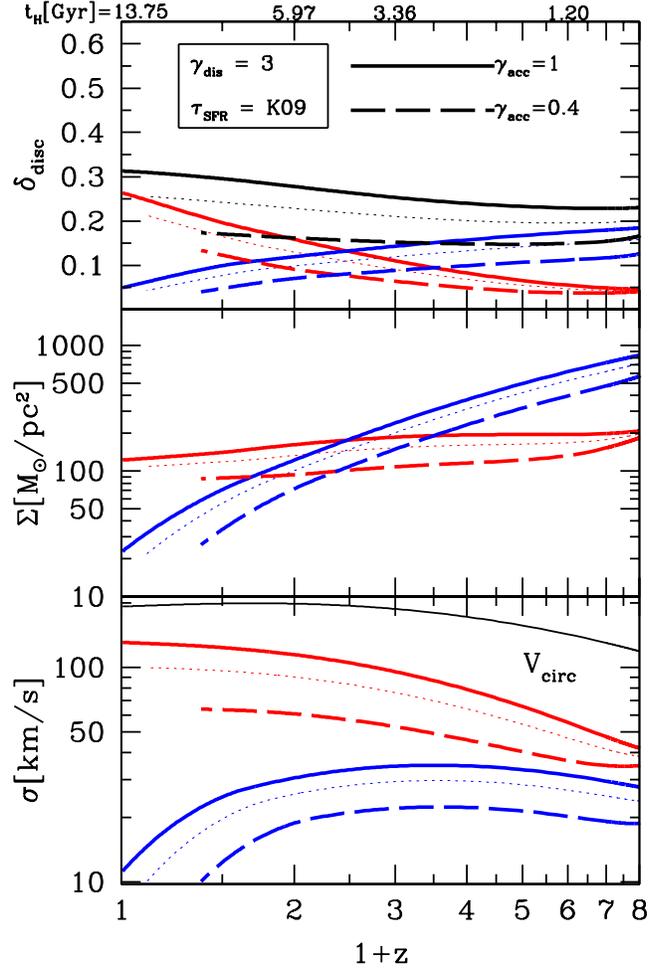,width=\hsize}}
\caption{
The effects of varying $\gamacc$ (\equ{gamacc}) in the range $0.4-1$
on the evolution of the two-component model.
Panels and curves are as in \fig{fig4} and \fig{fig5}.
The stabilization epoch is only weakly affected,
but the gas dispersion velocity is lower when the accretion onto the
disc is lower.
}
\label{fig:fig6}
\end{figure}
 
Here, we explore the effect varying $\gamacc$ (\equ{gamacc}),
the fraction of the average cosmological
accretion rate that joins the disc as gas,
using the extreme values $\gamacc = 0.4$ and $1$ about the
fiducial value of 0.7.
Fig.~\ref{fig:fig6} displays the results, showing that 
the stabilization time shifts from $z_{\rm stab} \simeq 0$ to $0.4$
when $\gamacc$ varies from 1 to 0.4.

Higher values of $\gamacc$ correspond to higher 
surface densities of the two components, because of the SFR dependence
on $\Sigg$. This dictates higher values of 
$\sigma$ for the two components in order to maintain $Q_{\rm 2c}=1$.
Both $\sigs/\sigg$
and $\Sigs/\Sigg$ are rather insensitive to
$\gamacc$, and so are, in particular, 
the time of turning star dominated and when $\sigg$ starts
dropping at $z \sim 1$.
The case of higher $\gamacc$ reaches $10 \kms$ later
because it has to drop from a higher value at earlier times,
$35\kms$ for $\gamacc=1$
compared to $20\kms$ in the case of $\gamacc=0.4$.

Note also in the top panel that a higher $\gamacc$ 
corresponds to a higher disc fraction, varying from 
$\deld\sim 0.17$ to $0.3$
when $\gamacc$ is varied from 0.4 to 1.

\begin{figure}
\centerline{\psfig{figure=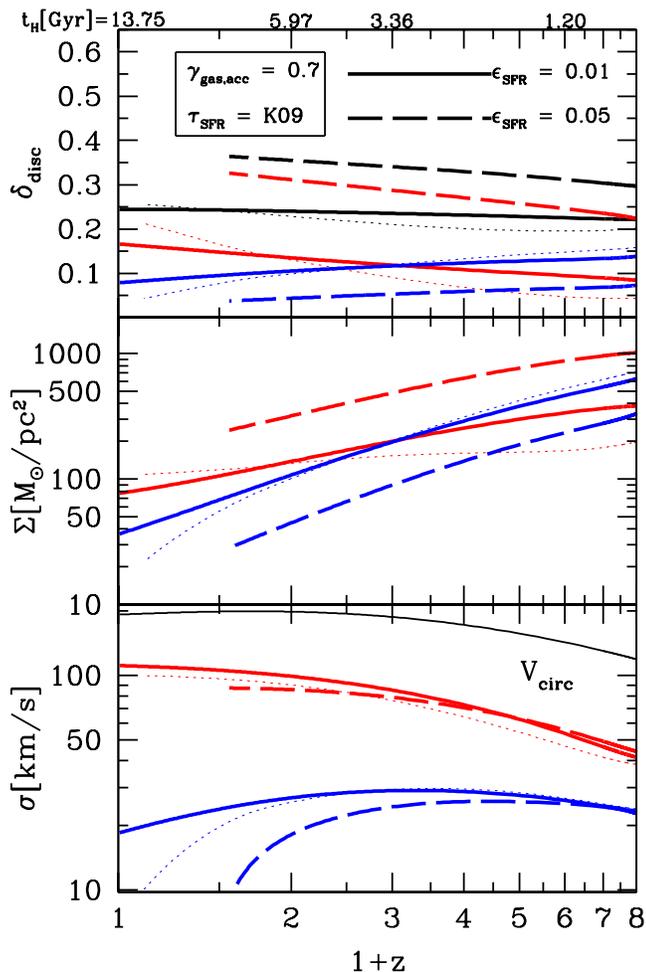,width=\hsize}}
\caption{
The effects of varying $\epssfr$ (\equ{sfr}) in the range $0.01-0.05$
on the evolution of the two-component model.
Panels and curves are as in \fig{fig4} and \fig{fig5}.
The stabilization epoch is rather sensitive to the SFR.
The case of very high SFR is unique in the sense that it is always stellar
dominated, and still, it maintains instability at high 
redshifts, to be 
stabilized only at late times.
}
\label{fig:fig7}
\end{figure}

\begin{figure}
\centerline{\psfig{figure=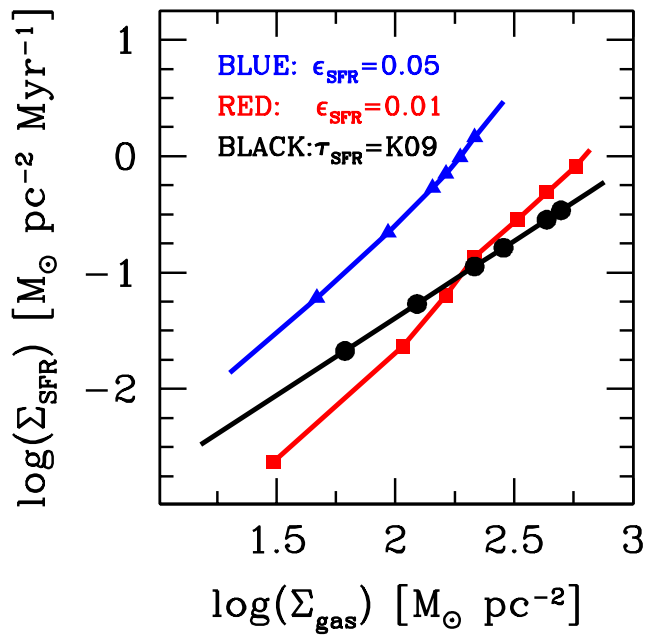,width=\hsize}}
\caption{
The SFR surface density as a function of gas surface density
for different SFR prescriptions:
the fiducial K09 prescription (black line, \equ{K09}), and 
the two cases shown in \fig{fig7}, with 
$\epssfr = 0.01$ (red line) and 0.05 (blue line). 
The points refer from right to left to $z=6,5,4,3,2,1$,
with the $\epssfr = 0.01$ extending to $z=0$. 
The SFR in the $\epssfr = 0.01$ case is higher than the fiducial model at 
early times, and lower at low redshifts.
}
\label{fig:fig7b}
\end{figure}

\subsection{Star Formation}
\label{sec:sfr}

We next study the effect of varying the star formation efficiency per free-fall
time, $\epssfr$ (\equ{sfr}), 
in comparison with our fiducial star-formation 
law of K09 \equ{K09}.
Fig~\ref{fig:fig7} shows the results for the cases with
$\epssfr = 0.01$ and $0.05$.

In order to help us interpret these results, we compare
in Fig.~10 the star formation laws, in terms of $\Sigma_{\rm SFR}$ as a function
of $\Sigma_{\rm gas}$, for the three models studied here.
We see that prior to $z \simeq 3$, the $\Sigma_{\rm SFR}$ in the
$\epssfr=0.01$ model is indeed slightly higher than in the K09 model,
and it becomes lower after this time.
On the other hand, the $\epssfr=0.05$ model has a significantly higher
$\Sigma_{\rm SFR}$ than the two other models at all times.
We will see that this has an impact on the evolution of the instability.

The model with $\epssfr=0.01$ remains unstable all the way to $z=0$
with an almost constant disc fraction in a steady state
at $\deld \simeq 0.25$, namely ${\rm B/D}\simeq 1.4$,
not too different from the one-component model 
(\S~\ref{sec:onecomp} and DSC09).
The reason is that the low SFR at late times 
allows the disc to never become strongly star dominated --- it maintains $\Sigma_{\rm stars}/\Sigg < 2$ over a long period
of time, from $z \sim 2$ to $0$. This eliminates the factor that serves as
the main driver for stabilization in the fiducial case.
On the other hand, at very high redshift, $\Sigs$ is higher 
than in the fiducial case. This is because in the high $\Sigg$
regime the SFR with a constant $\epssfr$ is higher than with the K09
law (see Fig.~10). 

The case with very efficient SFR, $\epssfr=0.05$, behaves
very differently from all other models. 
While it stabilizes at $z \simeq 0.6$, somewhat
earlier than the fiducial model, the very efficient SFR makes this disc 
strongly star dominated at all times, with 
$\Sigs/\Sigg \simeq 2.8$ already at $z=7$,
growing to $\sim 10$ at later times.  
Despite the stellar dominance, the disc is unstable till after $z \sim 1$.
The low gas fraction involves low dissipation and therefore slow inflow rate
in the disc, making the disc fraction grow to above $\deld=0.35$.
The reason for the rather decline of $\sigg$ starting at $z \sim 2$ towards
stabilization at $z \simeq 0.6$ is subtle, as no other quantity involved 
shows an abrupt change near that time. 
One thing that does change though is the rate of decline of $\Sigg$,
which at $z \sim 2$ becomes steeper than that of $\Omega \propto (1+z)^{3/2}$.
When this happens, $Q_{\rm 2c}$ can be maintained at unity only
if $\sigg$ declines, and this decline is enhanced because $\Sigma_{\rm gas}$
is low.

\subsection{Outflows due to Stellar Feedback}
\label{sec:outflow}

\begin{figure}
\centerline{\psfig{figure=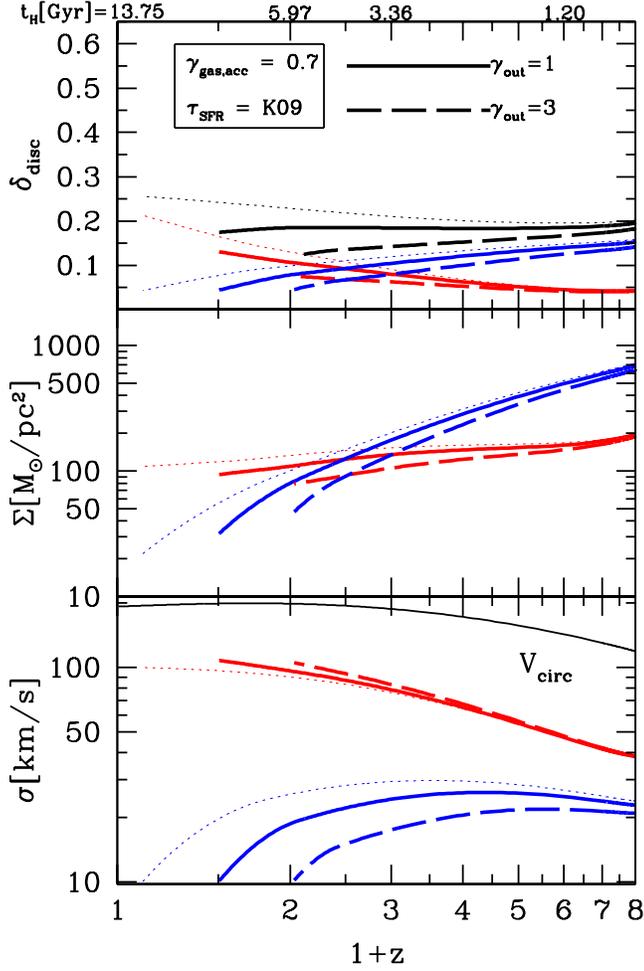,width=\hsize}}
\caption{
The effects of varying $\gamout$ (\equ{gamout}) in the range $0-3$
on the evolution of the two-component model.
Panels and curves are as in \fig{fig4} and \fig{fig5}.
Stronger outflows induce earlier stabilization, but still
limited to $z_{\rm stab} \sim 1$.
}
\label{fig:fig8}
\end{figure}

\Fig{fig8} shows the effects of strong and very strong outflows 
via $\gamout=1$ and 3 (\equ{gamout}).
We see that the stabilization epoch shifts from $z_{\rm stab} \simeq 0.15$
when outflows are ignored to $z_{\rm stab} \simeq 0.5$ and $1$ 
for $\gamout = 1$ and 3, respectively. 
This means that strong outflows could be the strongest driver towards early 
stabilization among the physical properties explored so far.
The gas removal from the disc due to strong outflows causes a steep decline
in $\Sigg$, making 
it drop below $100 \msun \pc^{-2}$ already prior to $z \simeq 1.5$
for $\gamout = 3$.
This translates via $Q_{\rm 2c}=1$ to a low gas velocity dispersion, dropping
below $20\kms$ already at $z \simeq 2.5$, from which the way to 
stabilization at $10\kms$ is rather short.
The disc fraction in the extreme outflow model becomes rather low, 
$\deld \simeq 0.1$, namely a bulge-dominated system.

With an outflow rate comparable to the SFR, $\gamout = 1$,
these effects are less drastic. 
While stabilization occurs at 
$z_{\rm stab} \approx 0.5$, the disc fraction is rather 
constant at about 0.2, and the disc at $z_{\rm stab}$ has $\Sigma_{\rm stars}/\Sigma_{\rm gas} \approx 4$ 
and $\sigma_{\rm stars}/\sigma_{\rm gas} \approx 10$, 
all not that different from the fiducial model in which outflows were
ignored.

\subsection{A Case of Extreme Stabilization}
\label{sec:extreme}

\begin{figure}
\centerline{\psfig{figure=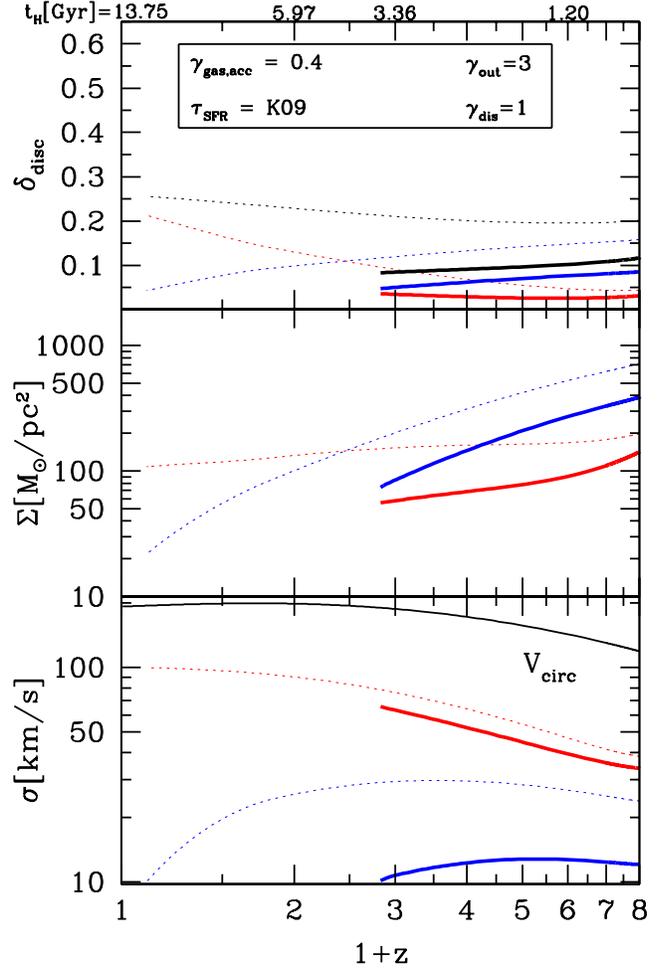,width=\hsize}}
\caption{
The effects of pushing the model parameters to extreme values 
favorable of early stabilization
on the evolution of the two-component model.
Panels and curves are as in \fig{fig4} and \fig{fig5}.
Even in this case, stabilization occurs only after $z = 2$.
}
\label{fig:fig9}
\end{figure}

In order to explore the robustness of the instability at high redshift, 
we present here a case in which the parameters are pushed to  
extreme limits within the sensible range, in an attempt to obtain
early stabilization.
This model has a very low fraction of gas accretion into the disc 
($\gamma_{\rm gas,acc}=0.4$), a high dissipation rate ($\gamdis=1$),
a high outflow rate ($\gamout=3$), and the K09 SFR law.
\Fig{fig9} demonstrates that even in such an extreme case the
disc remains unstable until $z \approx 2$. At the time of
stabilization, the surface density is not yet star dominated, 
but the low accretion rate and the strong outflows lead to a low total 
surface density, of $\Sigma \simeq 100\msun\pc^{-2}$ already by $z \simeq 2$.
The disc fraction is low and roughly constant at $\delta \simeq 0.1$, i.e.,
the system is bulge dominated with ${\rm B/D}\approx 5$. 
Thus, the disc stabilization in this case could be interpreted as driven
by ``morphological quenching" due the massive bulge (Martig et al. 2009)
rather than by the transition to a star-dominated disc.
We conclude that stabilization before $z \sim 2$ is very unlikely ---
it may be achieved only with more extreme outflows or in galaxies where the
accretion is severely suppressed compared to the cosmological average.

\section{Discussion and Conclusions}
\label{sec:concl}

We have studied the cosmological evolution of gravitationally unstable galactic
discs. To this aim, we have developed an analytic model that evolves in
time the properties of a galactic disc as a sequence of quasi steady-state
configurations under the assumption of self-regulated two-component
gravitational disc instability.
The disc is assumed to be fed by fresh gas at a constant fraction of the
average cosmological accretion rate.  
The gravitational instability is associated with
mass inflow within the disc towards a central bulge. 
In the disc, gas continuously turns into stars according
to a given star formation law. 
This typically leads to a characteristic pattern of evolution from
a gas-dominated disc at high redshifts to a star-dominated disc at low
redshifts, which is the main driver of instability at high redshifts
and the main reason for stabilization after $z \sim 1$.

The model is based on three key constraints, namely mass and energy
conservation and self-regulated instability. 
Mass conservation accounts for cosmolgical accretion onto the disc, mass
inflow within the disc into a central bulge, conversion of gas into stars,
and gas outflows. 
Energy conservation imposes that the gravitational energy gain due to the 
mass inflow down the potential gradient is driving velocity dispersion
in the disc gas and stars, compensating for the dissipation of the gas
turbulence on a dynamical timescale. 
The imposed self-regulated two-component instability then provides 
a connection between the velocity dispersions of gas and stars,
which allows us to solve the set of equations and propegate the system
in time.

While confirming that a single component disc, of gas or stars,
is likely to maintain the instability in a steady state till $z=0$ (DSC09),
we find here that a two-component disc, where gas continuously turns into
stars,
tends to (a) be unstable at $z>1$ and (b) stabilize at $z \sim 0-1$.
The instability at high redshift is typically driven by the high surface
density of gas and ``cold" stars, which is due to the high density of the
universe and the high cosmological accretion rate.
The stability at low redshift is typically driven by the steallar dominance,
which results from the declining cosmological accretion rate, 
the decline of $\Sigma$ because of the cosmological expansion,
the continuous convertion of gas to stars, the outflows of gas, 
and the effects of inflows in
the disc on disc depletion and bulge growth.
The low gas surface density and its fraction in the disc forces the gas
to a low velocity dispersion in order to balance the high stellar velocity
dispersion and keep $Q_{\rm 2c} \sim 1$. 

We study the impact of varying the assumed 
dissipation rate, fraction of gas accretion onto the
disc, star formation law, and outflow rate.
When the model parameters vary within a sensible range, the 
epoch of stabilization tend to change only in a limited range, typically
$0.1 <z<0.6$.
The inclusion of strong outflows plays a somewhat more significant role,
and it may stabilize the disc as early as $z \sim 1$ if
the mass removal rate is three times the SFR.
An extreme model, in which the parameters are all pushed to their most
favorable values for early stabilization, is capable of shifting the 
stabilization to an earlier epoch, but only to $z \sim 2$.
Prior to $z \sim 2$, it is difficult to avoid violent disc instability.

In a companion paper (Genel, Dekel \& Cacciato 2011), we explore the
alternative possibility that the disc turbulence is driven by the 
cosmological in-streaming in a non-self-regulated manner (as in DCS09). 
We find that also in this case the discs tend to evolve from instability     
to stability, and that $\sigma/V_{\rm circ}$ could be 
in the ballpark of the observed values at $z \sim 2$, with no need to 
appeal to feedback effects. However, in this case $\sigma/V_{\rm circ}$ 
is predicted to be independent of the actual value of the accretion rate, 
thus not reproducing the observed decline with time. 
When we do impose self-regulation at $Q \sim 1$, and introduce a duty
cycle for the instability, we find a better agreement with the observational
trends.

The current analysis is simplified in several ways that justify a few 
cautionary notes. First, the baryonic accretion is assumed to be at the average cosmological 
rate, while one expects significant variations in the accretion rate along the
history of every galaxy and between different galaxies as a function of their
environment. Since the accretion rate is the main driver of disc instability
and the associated phenomena, these variations may have important effects
on the evolution. The evolution of instability under realistic accretion 
histories will be addressed in a follow-up paper. 

Second,
in order to solve the set of equations, we had to make assumptions
concerning the stellar inflow rate in the disc compared to the gas inflow rate,
and especialy how the gravitational energy gain by the inflow is distributed 
between stirring up turbulence in the gas and raising the velocity disopersion 
of the stars.
Our assumptions boils down to assuming that 
(a) the mass inflow rate per unit mass, or the drift velocity, 
is the same for gas and stars,
and (b) that the energy deposited per unit mass in the disc
is the same for gas and stars.
The participation of stars in the inflow within the disc is obvious from
the fact that they dominate the giant clumps that are known from analytic estimates
(DSC09 and references therein) and simulations (Ceverino et al. 2011 and
references therein) to migrate inward on an orbital timescale as a result of 
torques, inter-clump interactions, and dynamical friction. 
However, the validity of the assumptions adopted here for the exact behavior of
the stars compared to the gas is still uncertain.
We note, for example, that Forbes et al. (in preparation), in a study of a similar problem,
are adopting a different assumption concerning the stellar migration rate
(M. Krumholz and J. Forbes, private communication).
The inflow rate in the two disc components, the migration of clumps compared 
to the inflow of disc mass outside the clumps, and the evolution of velocity
dispersion in the two components in the context of gravitational instability 
are all being studied in detail using high-resolution cosmological 
simulations (Cacciato et al. in preparation).

Third, we focused here on massive galaxies, comparable in mass to the Milky 
Way and to the observed big clumpy discs at $z \sim 2$ (Genzel et al. 2006;
2008; Elmegreen \& Elmegreen 2005).
However, one should be aware of the possibility that the evolution of 
instability may be different in less massive galaxies.
On one hand, at a given redshift, less massive galaxies tend to have a similar
$\Omega$ but a lower total surface density, which would require lower velocity
dispersions for maintaining $Q_{\rm 2c} \sim 1$, thus leading to earlier
stabilization. On the other hand, if the less massive galaxies are more gas
rich, as observed, perhaps due to different effects of star formation and
feedback, then stabilization is expected later in less massive galaxies.
There are indications for such a ``downsizing" in instability in observations
(Elmegreen \& Elmegreen 2005) and in simulations (Bournaud et al.~2011).
The mass dependence will be addressed in an upcoming paper.

Despite these limitations, our analysis indicates that phases of violent 
disc instability are a solid prediction
at $z > 1$, and that the typical discs are likely to stabilize prior to 
the present epoch, mostly due to the growing dominance of ``hot" stars 
at late times. The predictions of our analytic model are to be compared to those from high-resolution
hydro-cosmological simulations.

\section*{Acknowledgments}
We acknowledge stimulating discussions with Frederic Bournaud,  
Andi Burkert, Daniel Ceverino, John Forbes and Mark Krumholz.
This work has been supported by the
ISF through grant 6/08, by GIF through grant
G-1052-104.7/2009, by a DIP grant,
and by an NSF grant AST-1010033 at UCSC.
MC has been supported at HU by a Minerva fellowship (Max-Planck Gesellschaft). 
MC acknowledges the hospitality of the Dublin Institute for Advanced Studies 
and, together with AD, that of the Astronomy Department at UCSC.

\bibliography{flows}

\label{lastpage}
\end{document}